\documentclass{JHEP3} 

\usepackage{epsfig,multicol}
\usepackage{amsmath, amssymb}
\usepackage{epic}
\usepackage{concmath, palatino}
\usepackage{slashed}

\newcommand\fverb{\setbox\pippobox=\hbox\bgroup\verb}
\newcommand\fverbdo{\egroup\medskip\noindent%
			\fbox{\unhbox\pippobox}\ }
\newcommand\fverbit{\egroup\item[\fbox{\unhbox\pippobox}]}
\newbox\pippobox

\newcommand{\be}{\begin{equation}} 
\newcommand{\ee}{\end{equation}}

\newcommand{\ba}{\begin{eqnarray}}
\newcommand{\ea}{\end{eqnarray}}

\newcommand{\ovl}{\overline}
\newcommand{\mk}{\mathfrak}

\newcommand{\ads}{AdS_5\times S^5}

\title{Three loop anomalous dimensions of twist-3 gauge operators in ${\cal N}=4$ SYM}

\author{Matteo Beccaria\\
  Dipartimento di Fisica, Universita' del Salento,
  Via Arnesano, 73100 Lecce\\
  INFN, Sezione di Lecce\\
  E-mail: \email{matteo.beccaria@le.infn.it}}

\preprint{}

\abstract{
We propose a closed expression for the three loop anomalous dimension of a class of twist-3 operators
built with gauge fields and covariant derivatives. To this aim, we solve the long-range Bethe Ansatz 
equations at finite spin and provide a consistent analytical formula obtained assuming maximal transcendentality violation
as suggested by the known one-loop anomalous dimension. The final result reproduces the universal cusp anomalous dimension
and obeys recursion relations inspired by the principle of reciprocity invariance.
}

\begin{document} 

\section{Introduction}
\label{sec:Intro}

The maximal supersymmetric ${\cal N}=4$ super Yang-Mills theory plays a special role in the context of AdS/CFT duality~\cite{Maldacena:1997re}.
At the perturbative level, it is a finite superconformal theory with non trivial quantum properties of its gauge invariant
composite operators. The renormalization mixing can be seen as a dynamical system in the space of local composite operators
where the evolution time is the renormalization scale as first discovered in the QCD context (see~\cite{Belitsky:2004cz} for a review).
Remarkably, this system is integrable and can be related to a discrete universal $\mk{psu}(2,2|4)$ invariant superspin chain~\cite{Beisert:2004ry}.

It is commonly accepted that integrability survives at all loops. This important fact emerges from the Bethe Ansatz treatment of 
simple small rank subsectors. For the full theory, long-range (asymptotic) Bethe equations have been proposed~\cite{Beisert:2005fw}. 
They pass several consistency checks related to the integrability structure of the gauge theory, but also to the assumed duality with 
$\ads$ superstring. The full Bethe equations have a rather intriguing 
loop-deformed structure still to be completely understood from the point of view of perturbative deformations of integrable 
systems~\cite{Zabrodin:2007rq}.

Of course, with a very pragmatical attitude, integrability of ${\cal N}=4$ SYM can be exploited as a 
tool for the calculation of multi-loop anomalous dimensions. This is not as good as it could appear since solutions of Bethe equations cannot be found in 
analytical form, except for very special states. 
However, it turns out that certain operators, typically singlets of all discrete symmetries, enjoy a very remarkable property.
Their anomalous dimension is obtained as a weak-coupling perturbative series with {\em rational} coefficients (see~\cite{Genovese:2005ue}
for related comments).

This fact is  not a minor technical point. Indeed, one can try to resum these series in closed form 
in terms of the characteristic combinations of  harmonic sums which are a common result of multi-loop quantum field theory calculations.
The connection between Bethe equations and harmonic sums remains far from being 
understood, but is nevertheless a powerful heuristic tool to obtain anomalous dimensions for finite spin operators. 
Notice also that closed spin-dependent expressions are of paramount importance since they open the way to BFKL physics analyses~\cite{Kotikov:2007cy} as well as 
subtle QCD-like hidden properties like reciprocity relations~\cite{Basso:2006nk,Dokshitzer:2006nm,Beccaria:2007bb}.

As an important example, one can consider twist-2 operators in the bosonic $\mk{sl}(2)$ subsector of ${\cal N}=4$ SYM. 
For these operators,  Kotikov, Lipatov, Onishchenko and Velizhanin (KLOV) 
formulated a maximum transcendentality principle and predicted the three loop finite spin anomalous dimension 
from the QCD result~\cite{Kotikov:2004er,klov}. Later, those formulas have been checked against the rational predictions from Bethe Ansatz equations with full agreement,
although without a direct derivation~\cite{Staudacher:2004tk}.

A genuine new prediction has been recently obtained by considering similar twist-3 bosonic operators. The four loop finite spin 
anomalous dimension has been predicted by resorting to a weakly generalized KLOV principle~\cite{Beccaria:2007cn,Kotikov:2007cy}. Later, the same approach 
has been applied to twist-3 gaugino operators and the identification of rational anomalous dimensions has allowed to conjecture 
a relation with the twist-2 case proved rigorously at the three loop level~\cite{Beccaria:2007vh}.

The twist-3 case is quite interesting since operators built with scalars, gauginos or gauge fields are not related by supersymmetry.
This has to be compared with the twist-2 case where all channels are in a single supermultiplet~\cite{Belitsky:2003sh}.

In this paper, we continue the analysis of twist-3 operators by studying a purely gluonic sector. After a review of relevant known results, 
we  illustrate how the $\mk{psu}(2,2|4)$ supermultiplet structure in twist-3 can be exploited to identify the correct superconformal primary describing the 
gauge sector. At one-loop, this sector is described by the $XXX_{-3/2}$ closed spin chain. At higher orders, we solve perturbatively the 
long range Bethe equations. Applying the above strategy, we resum in closed form the anomalous dimensions up to three loops. 
To this aim, we propose a weak KLOV principle with subdominant transcendentality contributions.

Our three loop result reproduces in the large spin $N$ limit 
the universal twist-independent cusp anomalous dimension, {\em a.k.a} scaling function. Moreover, 
the various subleading corrections in an expansions at large $N$ are shown to obey generalized Moch-Vermaseren-Vogt 
relations~\cite{Moch:2004pa,Vogt:2004mw}
as suggested by general QCD arguments in~\cite{Basso:2006nk,Dokshitzer:2006nm,Beccaria:2007bb}.

\section{Quasi-partonic operators and planar integrability}

In this Section, we briefly recall the definition of a special class of QCD composite operators which 
are quite relevant in phenomenological applications, the so-called {\em quasi-partonic} operators~\cite{Belitsky:2004cz}.
They are defined with no special reference to the possible underlying supersymmetry.
As we shall discuss, quasi-partonic operators with maximal helicity have quite special properties from the point of 
view of both integrability and renormalization mixing. For this reason, they are a convenient bridge between the QCD language 
and that of the integrable ${\cal N}=4$ SYM theory and will be the starting point of our analysis in the next Sections.

\bigskip
We introduce light-cone coordinates by choosing two independent light-like 4-vectors $n^\mu$ and $\ovl n^\mu$ with
\be
n^2 = \ovl n^2 = 0,\qquad n\cdot\ovl n = 1,
\ee
and decompose a  generic 4-vector $V^\mu$ according to the relations
\ba
V^\mu &=& V_-\,n^\mu + V_+\,\ovl n^\mu+V_\perp^\mu,\qquad (V_\perp\cdot n = V_\perp\cdot\ovl n = 0). \\
V_+ &=& V\cdot n,\qquad V_- = V\cdot \ovl n,\qquad V_\perp^\mu = g^{\mu\nu}_{\perp\perp}\,V_\nu, \\
g_{\mu\nu}^\perp &=& g_{\mu\nu}-n_\mu\,\ovl n_\nu-\ovl n_\mu\,n_\nu.
\ea
A convenient choice is as usual $n^\mu = (1, 0, 0, 1)$, $\ovl n^\mu = (1, 0, 0, -1)$.

\bigskip
The relevance of light-cone is that in high energy QCD applications, one is often lead to consider 
operators with quark fields along the ``$-$'' ray and gauge links assuring gauge invariance. 
A typical 2-quark operator is then
\be
{\cal O}(z_1, z_2) = \ovl\psi(z_1\,n)\,\slashed{n} \,P\,e^{i\,g\,\int_{z_2}^{z_1} d\ell\, A_+(u\,\ell)}\,\psi(z_2\,n).
\ee
Understanding the gauge links, these non-local operators can be expanded in local operators with increasing spin
($D_\mu$ is the covariant derivative)
\be
{\cal O}(-z, z) = \sum_N\frac{(2\,z)^N}{N!}\,\ovl\psi(0)\,\slashed{n}\,\stackrel{\leftrightarrow}{D_+^N}\,\psi(0),\qquad D_+ = D\cdot n.
\ee

In the conformal limit, the ``$-$'' ray is left invariant by a $SL(2)$ collinear subgroup of the conformal group, generated
by translations and dilatations along the ray, and rotations in the $(+,-)$ plane. $SL(2)$ primary fields have 
definite scaling dimension $d$ and collinear spin $s$ defined by (here $D$ and $\Sigma_{\mu\nu}$ are the dilatation and  
Lorentz spin generators)
\be
D\, \Phi = d\, \Phi, \qquad \Sigma_{+-}\, \Phi = s\, \Phi.
\ee
The so-called {\em good fields} are special ($SL(2)$ primary) 
components of the elementary scalars $\varphi$ (in supersymmetric theories), Weyl fermions $\lambda_\alpha$
and field strength $F_{\mu\nu}$ with minimal collinear twist $t=d-s=1$. They can be shown to be
$\varphi$, $\lambda_+$ (and the conjugate) and $F^{+\mu}_{\ \ \, \perp}$.
Composite operators built with good fields are called {\em quasi-partonic} since they correspond to 
physical degrees of freedom as is clear in the light cone gauge. For quasi-partonic operators, the number of good fields equals the twist.
At one-loop, quasi-partonic operators with fixed twist are a closed set under renormalization mixing.

In the light-cone gauge, $A\cdot n = A_+ = 0$, the gauge links are absent and the physical fields are
 $\varphi$, $\lambda_+$ and $A^\mu_\perp$. Notice that the transverse components of the gauge field are 
non-locally related to good fields, for instance
\be
A = \frac{1}{\sqrt{2}}(A^x+i\, A^y) = \frac{1}{\sqrt 2}\partial_+^{-1}(F^{+ x}-i\, \widetilde{F}^{+ x}) =  \frac{i}{\sqrt 2}\partial_+^{-1}(F^{+ y}-i\, \widetilde{F}^{+ y}).
\ee

\bigskip
Quasi-partonic operators are a convenient starting point to discuss the emergence of integrability structures
in QCD, as well as in theories with ${\cal N}>0$ supersymmetries.
In the planar (multicolor) limit, the one-loop anomalous dimensions of various quasi-partonic QCD operators 
are computed by the spectrum of the integrable $XXX$ spin chain with sites transforming in 
negative spin (infinite dimensional) $\mk{sl}(2)$ representations.
Three quarks baryon operators with maximal helicity $3/2$ and leading twist-3 have 
anomalous dimensions associated with the closed $XXX_{-1}$ chain~\cite{Braun:1998id,Braun:1999te}.
Twist-3 quark-gluon chiral-odd operators lead to more complicated but still 
integrable open $XXX$ chains~\cite{Belitsky:1999qh,Belitsky:1999ru,Belitsky:1999hf,Derkachov:1999ze}.
Twist-3 maximal helicity gluon operators have anomalous dimensions described by the closed 
$XXX_{-\frac{3}{2}}$ chain~\cite{Belitsky:1999bf}.
The integrability of three gaugino operators survives at two loops in  ${\cal N}=1,2,4$ extended 
theories~\cite{Belitsky:2005bu}. The same is true for three scalar operators studied at ${\cal N}=2,4$
in~\cite{Belitsky:2006av}. In this case, the relevant one-loop integrable spin chain is $XXX_{-1/2}$.

In the maximally supersymmetric ${\cal N}=4$ theory, the integrability properties of various subsectors
(including quasi-partonic operators) are deeply intertwined with the large amount of supersymmetry.
In general, the $\mk{psu}(2,2|4)$ supermultiplet structure connects twist sectors and channels which are
completely unrelated in QCD. In the next Section, we shall review what is known about the anomalous dimensions
of twist-2 and -3 quasi-partonic operators with maximal helicity in ${\cal N}=4$ at one-loop and 
beyond. This preliminary discussion will set the stage for the investigation of a still unexplored (nice) 
piece of the theory which are purely twist-3 gluonic operators at more than one-loop.

\section{A variety of one and higher loop anomalous dimensions}

In the notation of~\cite{Belitsky:2006en}, let us consider a single-trace maximal helicity quasi-partonic operator
\be
{\cal O}_{N, L}(0) = \sum_{n_1+\cdots n_L=N} a_{n_1,\dots n_L}\,\mbox{Tr}\left\{D_+^{n_1} X(0)\cdots D_+^{n_L} 
X(0)\right\},\ \ n_i\in\mathbb{N},
\ee
where $X(0)$ is a physical component of quantum fields with definite helicity in the 
underlying gauge theory ($\varphi, \lambda, A$), and $D_+$ is the light-cone projected covariant derivative. 
The coefficients $\{a_\mathbf{n}\}$ are such that ${\cal O}_{N, L}$ is a scaling field, eigenvector of the dilatation
operator. The total Lorentz spin is $N=n_1 + \cdots n_L$. The number of elementary fields equals the twist 
$L$, {\em i.e.} the classical dimension minus the Lorentz spin.

\medskip
At one-loop, the anomalous dimensions of the above operators can be found from the spectrum of a 
noncompact $\mathfrak{sl}(2)$ spin chain with $L$ sites. 
The elementary spin of the chain is related to the conformal spin $s$ of $X$
which is $s = \frac{1}{2}, 1, \frac{3}{2}$ when $X$ is a scalar, gaugino, or gauge field respectively.
The one-loop ground state energy, associated with the lowest anomalous dimension, can be found by solving the 
Baxter equation~\cite{Bax72}. For self-consistency, we recall the well-known basic equations.
One introduces the Baxter function $Q(u)$  satisfying the second-order finite-difference equation
\be
(u + i\,s)^L\, Q (u + i) + (u - i\,s)^L\, Q(u - i) = t_L(u)\, Q(u),
\ee
where $t_L(u)$ is a polynomial in $u$ of degree $L$ with coefficients given by conserved charges $q_i$
\be
t_L(u) = 2\, u^L + q_2\, u^{L-2} + \ldots + q_L.
\ee
The lowest integral of motion is 
\be
q_2 = -(N+L\,s) (N+L\,s - 1) + L\, s\,(s-1),
\ee
with $N=0,1,\ldots$. If the Baxter function is assumed to be a polynomial of degree $N$
\be
Q(u) \sim \prod_{k=1}^N (u-u_k),
\ee
one immediately checks that the Baxter equation implies the Bethe equations
for the $XXX_{-s}$ chain
\be
\left(\frac{u_k+i\,s}{u_k-i\,s}\right)^L=\mathop{\prod_{j=1}^N}_{j\neq k}
\frac{u_k-u_j-i}{u_k-u_j+i}\qquad k=1, \dots, N.
\ee
Solving the Baxter equation supplemented by the polynomiality constraint one easily
obtains quantized values of the charges $q_3,\ldots,q_L$ and evaluates the
corresponding energy and quasimomentum from
\be
\varepsilon = i\left(\ln Q(i\,s)\right)'-i\left(\ln Q(-i\,s)\right)'\,,\qquad e^{i\theta} =
\frac{Q(i\,s)}{Q(-i\,s)}\,.
\ee
As usual, the cyclic symmetry of the single-trace operators requires $e^{i\theta}=1$. The one-loop anomalous
dimension of Wilson operators are related to the chain energies $\varepsilon$ by
\be
\Delta \gamma(N) = g^2\,\varepsilon(N) + {\cal O}(g^4),
\ee
where $g^2 = \lambda/(8\,\pi^2) = g_{\rm YM}^2\,N_c/(8\,\pi^2)$ is the scaled 't Hooft coupling, 
fixed in the planar $N_c\to\infty$ limit.
In the above expressions, $\Delta\gamma(N) = \gamma(N)-\gamma(0)$ 
is the subtracted anomalous dimension vanishing at zero spin $N=0$.

\subsection{Twist-2, SUSY universality and the KLOV principle}

As an example and for illustrative purposes, let us briefly review what happens in twist-2. Solving the 
Baxter equation in the three sectors $s = 1/2, 1, 3/2$, {\em i.e.} for the scalar ($\varphi$), 
gaugino ($\lambda$) and vector ($A$) channels, one immediately recovers the known one-loop formulae
\ba
\Delta\gamma_{L=2}^\varphi(N) &=& 4\,S_1(N), \nonumber \\
\Delta\gamma_{L=2}^\lambda(N) &=& 4\,S_1(N+1)-4, \\
\Delta\gamma_{L=2}^A(N) &=& 4\,S_1(N+2)-6,
\ea
where $S_1(N) = \sum_{n=1}^N \frac{1}{n}$ is the $N$-th harmonic number. 

These results express the well-known fact that all twist-2 quasipartonic operators are in the same SUSY multiplet
and their anomalous dimension is expressed by a universal function with shifted arguments~\cite{Belitsky:2003sh}.
Taking into account $\gamma(0)$ one has 
\be
\gamma_{L=2}^\varphi(N) = \gamma_{\rm univ}(N),\qquad
\gamma_{L=2}^\lambda(N) = \gamma_{\rm univ}(N+1), \qquad
\gamma_{L=2}^A(N) = \gamma_{\rm univ}(N+2). 
\ee
These relations are a consequence of the unbroken superconformal symmetry and are expected to hold at
all orders. The higher order (three loop) corrections to $\gamma_{\rm univ}$ are available due to the deep insight of 
Kotikov, Lipatov, Onishchenko and Velizhanin (KLOV)~\cite{Kotikov:2004er} 
(see also~\cite{klov}). Their prediction is based on what is now universally known as the {\em maximum transcendentality or 
KLOV principle}. 

Expanding the universal function $\gamma_{\rm univ}(N)$ according to 
\be
\gamma_{\rm univ}(N) = \sum_{n\ge 1}\gamma_{\rm univ}^{(n)}(N)\,g^{2\,n} ,
\ee
the explicit result is~\cite{Kotikov:2004er}
\ba
\label{eq:univ}
\gamma_{\rm univ}^{(1)}(N) &=& 4\, S_1\, ,  \\
\gamma_{\rm univ}^{(2)}(N) &=&-4\,\Big( S_{3} + S_{-3}  -
2\,S_{-2,1} + 2\,S_1\,\big(S_{2} + S_{-2}\big) \Big)\, ,  
\nonumber \\
\gamma_{\rm univ}^{(3)}(N) &=& -8 \Big( 2\,S_{-3}\,S_2 -S_5 -
2\,S_{-2}\,S_3 - 3\,S_{-5}  +24\,S_{-2,1,1,1}\nonumber\\
&&~~~~~~+ 
6\,\big(S_{-4,1} + S_{-3,2} + S_{-2,3}\big)
- 12\,\big(S_{-3,1,1} + S_{-2,1,2} + S_{-2,2,1}\big)\nonumber \\
&&~~~~~~-
\big(S_2 + 2\,S_1^2\big) 
\big( 3 \,S_{-3} + S_3 - 2\, S_{-2,1}\big)
- S_1\,\big(8\,S_{-4} + S_{-2}^2\nonumber \\
&&~~~~~~+ 
4\,S_2\,S_{-2} +
2\,S_2^2 + 3\,S_4 - 12\, S_{-3,1} - 10\, S_{-2,2} 
+ 16\, S_{-2,1,1}\big)
\Big)\, , \nonumber
\ea
with all harmonic sums evaluated at argument $N$ and nested sums defined recursively by 
\be
S_a(N) = \sum_{n=1}^N\frac{(\mbox{sign}\,a)^n}{n^a}, \qquad S_{a_1, a_2, \dots}(N) = \sum_{n=1}^N\frac{(\mbox{sign}\,a_1)^n}{n^{a_1}}\,S_{a_2, \dots}(n).
\ee
Expressions Eqs.~(\ref{eq:univ}) have been shown to be fully consistent with the long-range Bethe Ansatz equations
valid in the bosonic $\mk{sl}(2)$ sector~\cite{Staudacher:2004tk} by checking them at many values of the spin $N$. It must be 
emphasized that a direct proof of such expressions from the Bethe equations is still missing beyond one-loop.

At more than 3 loops, wrapping problems forbid to 
predict the anomalous dimension of twist-2 operators with finite spin~\cite{Kotikov:2007cy}.
Nevertheless, it is possible to predict from the Bethe Ansatz the all-loop expansion of the scaling function $f(g)$ defined by 
the large spin limit
\be
\label{eq:scaling}
\gamma_{\rm univ}(N) = f(g)\,\log\,N + {\cal O}(1).
\ee
At four loops, the analytical prediction reported in~\cite{Beisert:2006ez} is in full agreement with the alternative 
(more conventional) calculations of~\cite{Bern:2006ew,Cachazo:2006az}.

\subsection{Twist-3, Universality classes and KLOV breaking}

The same one-loop exercise at twist-3 gives, for even spin $N$ ( a necessary condition to select an unpaired ground state)
\ba
\label{eq:twist3}
\Delta\gamma_{L=3}^\varphi(N) &=& 4\,S_1\left(\frac{N}{2}\right), \nonumber \\
\Delta\gamma_{L=3}^\lambda(N) &=& 4\,S_1(N+2)-6, \\
\Delta\gamma_{L=3}^A(N) &=& 4\,S_1\left(\frac{N}{2}+1\right)-5+\frac{4}{N+4}.\nonumber
\ea
The result for the scalar channel is rather well understood and its four-loop correction has been 
recently computed in two independent papers~\cite{Beccaria:2007cn,Kotikov:2007cy}.

Apart from a constant shift related to $\gamma(0)$, 
the anomalous dimension in the gaugino channel is clearly related to the twist-2 universal anomalous dimension.
This remarkable degeneracy among different twist operators has been first studied in~\cite{Korchemsky:1994um,Derkachov:2002wz}.
Recently, it has been proved that supersymmetry implies (at least) the three loop relation~\cite{Beccaria:2007vh}
\be
\label{eq:conj}
\gamma_{L=3}^\lambda(N) = \gamma_{\rm univ}(N+2),\qquad N\in 2\mathbb{N}.
\ee
In the gauge sector, the above one-loop result is fully 
consistent with the analysys of maximal helicity 3 gluon operators in QCD~\cite{Belitsky:1999bf}. The dilatation operator
is integrable and its lowest eigenvalue is given by Eq.~(82) of~\cite{Belitsky:1999bf}:
\ba
\varepsilon &=& 2\,S_1\left(\frac{N}{2}+2\right)+2\,S_1\left(\frac{N}{2}+1\right)+4 = \\
&=& 4\,S_1\left(\frac{N}{2}+1\right)+\frac{4}{N+4}+4.
\ea
Apart from the constant, this is precisely the same combination appearing in $\gamma_{L=3}^A$. 
Also, at one-loop this prediction is expected to agree with the ${\cal N}=4$ result, thus fixing $\gamma(0)$.

\medskip
We remark that, already this very simple one-loop analysis, reveals the following non-trivial features
of the twist-3 case
\begin{enumerate}
\item In twist-3 there are various universality classes of anomalous dimensions as a consequence of a richer supermultiplet
structure.
\item The twist-2 universality class is inherited in the gaugino sector.
\item In the gauge sector, the KLOV principle is violated at least in strict sense.
\end{enumerate}
Items (1,2) are expected to hold at all orders being related to the superconformal symmetry. Concerning (3), it would 
be quite interesting to explore what happens beyond the one-loop level. This is a somewhat unexplored territory to which 
we devote the remaining part of this paper.

\section{Tripleton decomposition and twist-3 anomalous dimensions}

The elementary fields of ${\cal N}=4$ SYM are (in a chiral basis and omitting the fermion $\mathfrak{su}(4)$ index)
\be
\varphi,\quad, \lambda_\alpha, \quad\overline\lambda^{\dot\alpha}, \quad F_{\alpha\beta}, \quad \overline{F}_{\dot\alpha\dot\beta}.
\ee
Together with their derivatives, they belong to an irreducible representation of the superconformal algebra, 
the {\em singleton} $V_F$.

From the superconformal properties of tensor products $V_F^{\otimes L}$ we can understand several general features 
of the twist-3 anomalous dimensions.
The decomposition of  $V_F^{\otimes L}$ in irreducible superconformal representations is 
nicely discussed in~\cite{Beisert:2004di} exploiting 
the higher spin symmetry $\mathfrak{hs}(2,2|4)$ of the free theory.

As a warm-up, let us reconsider the twist-2 case. We label a superconformal multiplet ${\cal V}$ by the quantum numbers of 
its superconformal primary state
\be
{\cal V}^{\Delta, B}_{[j, \overline j][\lambda_1, \lambda_2, \lambda_3]},
\ee
where $\Delta$ is the conformal scaling dimension, $B$ the hypercharge~\cite{Intriligator:1998ig}, $[j, \overline j]$ the 
chiral and antichiral $\mathfrak{su}(2)\oplus\mathfrak{su}(2)$ Lorentz spins, and $[\lambda_1, \lambda_2, \lambda_3]$ the Dynkin labels
of the $\mathfrak{su}(4)$ $R$-symmetry algebra.

In standard notation, the symmetric product of two singleton representations decomposes as 
\be
(V_F\otimes V_F)_S = \bigoplus_{n=0}^\infty V_{2n},
\ee
where $V_n$ is an abbreviation for the semishort current multiplet
\be
V_n \equiv {\cal V}^{n, 0}_{[\frac{n}{2}-1, \frac{n}{2}-1][0,0,0]}.
\ee
The multiplets $V_{2n}$ contain, among others, all twist-2 operators with increasing spin. The anomalous dimension of the 
multiplet is proportional to the harmonic number $S_1(2\,n)$, at one-loop,  
and is nothing but the universal anomalous dimension $\gamma_{\rm univ}^{(1)}$.

The decomposition in twist-3 is much more complicated. Following~\cite{Beisert:2004di}, the maximally symmetric case reads
\be
(V_F\otimes V_F\otimes V_F)_S = \mathop{\bigoplus_{n=0}}_{k\in\mathbb{Z}}^\infty c_n\left[V_{2k, n}+V_{2k+1, n+3}\right],
\ee
where $c_n = 1+[n/6]-\delta_{n, 1\,{\rm mod}\, 6}$. The various modules appearing in the decomposition have quite different properties
and describe states in various subsectors. 

For even $n$, the {\em one-loop} lowest anomalous dimension in $V_{m,n}$ is associated with an unpaired state and has 
been guessed in~\cite{Beisert:2004di} to be 
\be
\gamma_{m, n} = \frac{\lambda}{8\pi^2}\left[
2\,S_1\left(\frac{m}{2}-\frac{1}{2}\right)
+2\,S_1\left(m+\frac{n}{2}\right)
+2\,S_1\left(\frac{m}{2}+\frac{n}{2}\right)
-2\,S_1\left(-\frac{1}{2}\right)\right].
\ee
Taking $m=2$ and $n=N$ we find 
\be
\label{eq:oneloopmodule}
\gamma_{2, N} =  \frac{\lambda}{8\pi^2}\left[
2\,S_1\left(\frac{N}{2}+1\right)
+2\,S_1\left(\frac{N}{2}+2\right)+4\right],
\ee
where we have used the representation $S_1(N) = \gamma_E + \psi(N+1)$ in terms of the digamma function $\psi(z) = (\log\Gamma(z))'$
which satisfies
\be
\psi\left(\frac{2\,n+1}{2}\right) - \psi\left(\frac{2\,n-1}{2}\right) = \frac{2}{2\,n-1},\qquad n\in \mathbb{N}.
\ee
The expression Eq.~(\ref{eq:oneloopmodule}) reproduces the three gluon one-loop anomalous dimension in Eq.~(\ref{eq:twist3}). The reason of the agreement is that 
the module $V_{2,N}$ describes states in a $\mk{su}(1,2)$ subsector which are covariant derivatives of the self-dual field strength~\cite{Beisert:2004di}.
Indeed, the associated superconformal primary state $\Phi_{\rm gauge}$
is built with $L=3$ fields and its quantum numbers can be read from the relation
\be
V_{2, N} = {\cal V}^{N+4, 1}_{[\frac{N}{2}+1, \frac{N}{2}][0,0,0]}.
\ee
Using the oscillator representation discussed in~\cite{Beisert:2003jj}, one checks that, after a Lorentz rotation, $\Phi_{\rm gauge}$ has the schematic form 
\be
\Phi_{\rm gauge} = \mbox{Tr}(D_{11}^N\, \lambda_{11}\,\lambda_{11}\, \varphi_{34} + \cdots),
\ee
where $\lambda_{\alpha\ a}$, $\alpha=1,2$, $a=1, \dots, 4$ are the Weyl fermions
and $\varphi_{ab} = -\varphi_{ba}$, $a,b = 1, \dots, 4$, the six real scalars transforming in the  {\bf 4} and {\bf 6} of $\mk{su}(4)$ 
respectively. Applying the four supersymmetry charges $Q^a_\alpha$ with $a=1,2,3,4$ and $\alpha=1$, we reach states of the form 
\be
Q^1_1\, Q^2_1\, Q^3_1\, Q^4_1\, \Phi_{\rm gauge} = \mbox{Tr}(D_{11}^N\, F_{11}\,F_{11}\,F_{11} + \cdots) + \cdots.
\ee
In light-cone coordinates, $D_{11} = D_+$ and $F_{11}$ is the holomorphic combination of $F^{+\mu}_{\ \ \ \ \perp}$ with definite
helicity.
Thus, if we want to discuss the multiloop anomalous dimension of twist-3 maximal helicity quasi-partonic operators, the correct 
context in the full $\mk{psu}(2,2|4)$ theory is precisely module $V_{2,N}$ for even $N$. 

As a check, we illustrate in the next Section the useful exercise of recovering the one-loop equivalence with the $XXX_{-\frac{3}{2}}$ chain. Next, we shall 
study the solution of the long-range Bethe equations for $\Phi_{\rm gauge}$ in order to obtain higher loop 
predictions.

\section{One-loop reduction to the $XXX_{-\frac{3}{2}}$ spin chain}

As discussed  in~\cite{Beisert:2003yb}, the Bethe Ansatz equations for the full ${\cal N}=4$ theory with
(complexified) algebra $\mk{sl}(4|4)$ must have a 
rather universal structure discussed in~\cite{Ogievetsky:1986hu} for bosonic symmetry algebras 
and in~\cite{Saleur:1999cx} in the fermionic case.

This general structure may be written as follows.
Suppose that the symmetry algebra has rank $r$. Let us look for a state associated with $K = K_1 + \cdots + K_r$ Bethe roots
denoted by $u_i$, $i=1, \dots, K$.
For each root we specify which of the $r$ simple roots is excited by
$k_j=1,\ldots,r$.
The Bethe equations can be written in the compact form 
\be
\left(\frac{u_j+\frac{i}{2}V_{k_j}}{u_j-\frac{i}{2}V_{k_j}}\right)^L =
\mathop{\prod_{\ell=1}^K}_{\ell\neq j} \frac{u_j-u_\ell+\frac{i}{2}M_{k_j,k_\ell}}{u_j-u_\ell-\frac{i}{2}M_{k_j,k_\ell}}.
\ee
Here, $M_{k\ell}$ is the Cartan matrix of the algebra and $V_k$ 
are the Dynkin labels of the spin representation carried by each site of the chain.
Furthermore, we still consider a cyclic spin chain with zero
total momentum and this gives the additional constraint
\be
1=\prod_{j=1}^K\frac{u_j+\frac{i}{2}V_{k_j}}{u_j-\frac{i}{2}V_{k_j}}.
\ee
The energy of a configuration of roots that
satisfies the Bethe equations and constraint is now given, apart from $R$-matrix ambiguities (encoded in the constant $c$
and the choice of sign)  by
\be
E=c L \pm \sum_{j=1}^K\left(\frac{i}{u_j+\frac{i}{2}V_{k_j}}-\frac{i}{u_j-\frac{i}{2}V_{k_j}}\right).
\ee

In the particular case of a $\mk{psu}(2,2|4)$-invariant theory, we need to specify the Cartan matrix,
determined by the Dynkin diagram which is not unique for superalgebras, as well as the Dynkin labels of the
spin representation corresponding to the singleton module $V_F$.
In the context of $N=4$ SYM, a convenient choice is the {\em Beauty} form 
\medskip
\be
\begin{minipage}{260pt}
\setlength{\unitlength}{1pt}%
\small\thicklines%
\begin{picture}(260,35)(-10,-10)
\put(  0,00){\circle{15}}%
\put(  0,10){\makebox(0,0)[b]{}}%
\put(  7,00){\line(1,0){26}}%
\put( 40,00){\circle{15}}%
\put( 40,15){\makebox(0,0)[b]{}}%
\put( 47,00){\line(1,0){26}}%
\put( 80,00){\circle{15}}%
\put( 80,15){\makebox(0,0)[b]{}}%
\put( 87,00){\line(1,0){26}}%
\put(120,00){\circle{15}}%
\put(120,15){\makebox(0,0)[b]{+1}}%
\put(127,00){\line(1,0){26}}%
\put(160,00){\circle{15}}%
\put(160,15){\makebox(0,0)[b]{}}%
\put(167,00){\line(1,0){26}}%
\put(200,00){\circle{15}}%
\put(200,15){\makebox(0,0)[b]{}}%
\put(207,00){\line(1,0){26}}%
\put(240,00){\circle{15}}%
\put(240,15){\makebox(0,0)[b]{}}%
\put( 35,-5){\line(1, 1){10}}%
\put( 35, 5){\line(1,-1){10}}%
\put(195,-5){\line(1, 1){10}}%
\put(195, 5){\line(1,-1){10}}%
\end{picture}
\end{minipage}
\ee

\medskip
\medskip
On top of the Dynkin diagram 
we have indicated the
Dynkin labels of the spin representation.
We write the Cartan matrix corresponding to this choice of Dynkin diagram
and the representation vector as%
\be
M=\left(\begin{array}{c|c|ccc|c|c}
-2&+1&  &  &  &  &   \\\hline   
+1&  &-1&  &  &  &   \\\hline
  &-1&+2&-1&  &  &   \\
  &  &-1&+2&-1&  &   \\
  &  &  &-1&+2&-1&   \\\hline
  &  &  &  &-1&  &+1 \\\hline
  &  &  &  &  &+1&-2
\end{array}\right),\qquad
V=\left(\begin{array}{r}
0\\\hline0\\\hline0\\1\\0\\\hline0\\\hline0
\end{array}\right).
\ee
The energy corresponding to a solution to the Bethe equations is 
\be
E = \sum_{j=1}^K\left(\frac{i}{u_j+\frac{i}{2}V_{k_j}}-\frac{i}{u_j-\frac{i}{2}V_{k_j}}\right) = 
\sum_{j=1}^K\frac{V_{k_j}}{u_j^2 + \frac{1}{4}\,V_{k_j}^2}.
\ee
Given the quantum number of the superconformal state we are interested in, one can compute the excitation numbers $K_i$
according to the detailed expressions reported in~\cite{Beisert:2003yb}.

\subsection{Duality Transformations}

It is convenient to define the following graphical notation to denote bosonic or fermionic simple roots of a (modified)
superalgebra of $\mk{sl}(n|m)$ type as follows. 

There is a single type of fermionic nodes
\be
(F)
\begin{minipage}{260pt}
\setlength{\unitlength}{1pt}
\small\thicklines
\begin{picture}(260,55)(-10,-30)
\put(50,-15){\makebox(0,0)[t]{$M_{j, j-1}=+1$}}
%
\dottedline{3}(88,0)(112,0)  
\put(120,00){\circle{15}}
\put(115,-5){\line(1, 1){10}}  
\put(115, 5){\line(1,-1){10}}  
\put(120,15){\makebox(0,0)[b]{$V_j$}} 
\put(120,-15){\makebox(0,0)[t]{$M_{j,j} = 0$}} 
\put(127,00){\line(1,0){26}} 
%
\put(200,-15){\makebox(0,0)[t]{$M_{j, j+1}=-1$}}
\end{picture}
\end{minipage}
\ee
Eventually, it can appear in flipped form 
\be
(F')
\begin{minipage}{260pt}
\setlength{\unitlength}{1pt}
\small\thicklines
\begin{picture}(260,55)(-10,-30)
\put(50,-15){\makebox(0,0)[t]{$M_{j, j-1}=-1$}}
\put( 87,00){\line(1,0){26}} 
\put(120,00){\circle{15}}
\put(115,-5){\line(1, 1){10}}  
\put(115, 5){\line(1,-1){10}}  
\put(120,15){\makebox(0,0)[b]{$V_j$}} 
\put(120,-15){\makebox(0,0)[t]{$M_{j,j} = 0$}} 
\dottedline{3}(128,0)(152,0) 
\put(200,-15){\makebox(0,0)[t]{$M_{j, j+1}=+1$}}
\end{picture}
\end{minipage}
\ee
There are two types of bosonic nodes
\be
(B)
\begin{minipage}{260pt}
\setlength{\unitlength}{1pt}
\small\thicklines
\begin{picture}(260,55)(-10,-30)
\put(50,-15){\makebox(0,0)[t]{$M_{j, j-1}=-1$}}
\put( 87,00){\line(1,0){26}} 
\put(120,00){\circle{15}}
\put(120,15){\makebox(0,0)[b]{$V_j$}} 
\put(120,-15){\makebox(0,0)[t]{$M_{j,j} = 2$}} 
\put(127,00){\line(1,0){26}} 
%
\put(200,-15){\makebox(0,0)[t]{$M_{j, j+1}=-1$}}
\end{picture}
\end{minipage}
\ee

\be
(\ovl B)
\begin{minipage}{260pt}
\setlength{\unitlength}{1pt}
\small\thicklines
\begin{picture}(260,55)(-10,-30)
\put(50,-15){\makebox(0,0)[t]{$M_{j, j-1}=+1$}}
%
\dottedline{3}(88,0)(112,0)  
\put(120,00){\circle{15}}
\put(120,15){\makebox(0,0)[b]{$V_j$}} 
\put(120,-15){\makebox(0,0)[t]{$M_{j,j} = -2$}} 
\dottedline{3}(128,0)(152,0) 
\put(200,-15){\makebox(0,0)[t]{$M_{j, j+1}=+1$}}
\end{picture}
\end{minipage}
\ee
In the  Beauty form, the Dynkin diagram of $\mk{psu}(2,2|4)$ is

\be
\begin{minipage}{260pt}
\setlength{\unitlength}{1pt}
\small\thicklines
\begin{picture}(260,55)(-10,-30)
\dottedline{3}(-32,0)(-10,0)  
\put(  0,00){\circle{15}}
\dottedline{3}(8,0)(32,0)    
\put( 40,00){\circle{15}}     
\put( 35,-5){\line(1, 1){10}}  
\put( 35, 5){\line(1,-1){10}}  
\put( 47,00){\line(1,0){26}} 
\put( 80,00){\circle{15}}
\put( 87,00){\line(1,0){26}} 
\put(120,00){\circle{15}}
\put(120,15){\makebox(0,0)[b]{$+1$}} 
\put(127,00){\line(1,0){26}} 
\put(160,00){\circle{15}}
\put(167,00){\line(1,0){26}} 
\put(200,00){\circle{15}}
\put(195,-5){\line(1, 1){10}} 
\put(195, 5){\line(1,-1){10}} 
\dottedline{3}(208,0)(232,0) 
\put(240,00){\circle{15}}
\dottedline{3}(250,0)(270,0) 
\end{picture}
\end{minipage}
\ee
where we have added external lines required to assess nodes 1, 7 as bosonic.

\bigskip
Now, let us consider a fermionic node with $K^0$ Bethe roots and neighbouring nodes
with $K^\pm$ Bethe roots, the sign being that of the associated off diagonal Cartan matrix element
\be
\begin{minipage}{260pt}
\setlength{\unitlength}{1pt}
\small\thicklines
\begin{picture}(260,55)(-10,-30)
\put(50,-15){\makebox(0,0)[t]{$K^+$}}
%
\dottedline{3}(88,0)(112,0)  
\put(120,00){\circle{15}}
\put(115,-5){\line(1, 1){10}}  
\put(115, 5){\line(1,-1){10}}  
\put(120,15){\makebox(0,0)[b]{$V^0$}} 
\put(120,-15){\makebox(0,0)[t]{$K^0$}} 
\put(127,00){\line(1,0){26}} 
%
\put(200,-15){\makebox(0,0)[t]{$K^-$}}
\end{picture}
\end{minipage}
\ee
The neighbouring nodes can be {\bf bosonic or fermionic}.
It is possible to prove duality relations that allows to write the Bethe equations in equivalent forms
related to modified Dynkin diagrams. Each dualization flips the two lines entering a fermionic node and changes its
excitation number $K^0\to \widetilde{K^0}$ as well as the weights of the three involved nodes. The precise rules
are proved and discussed in~\cite{Beisert:2005di}. They can be summarized in the following two duality transformations

\medskip
{\bf Dualization I}: $V^0\neq 0$, \ \ \ $\widetilde{K^0} = L+K^++K^--K^0-1$
\be
\begin{minipage}{260pt}
\setlength{\unitlength}{1pt}
\small\thicklines
\begin{picture}(260,55)(-10,-30)
\put( 80,00){\circle{15}}
\put( 80,15){\makebox(0,0)[b]{$V^+$}} 
\put( 80,-15){\makebox(0,0)[t]{$K^+$}}  
\dottedline{3}(88,0)(112,0)  
\put(120,00){\circle{15}}
\put(115,-5){\line(1, 1){10}}  
\put(115, 5){\line(1,-1){10}}  
\put(120,15){\makebox(0,0)[b]{$V^0$}} 
\put(120,-15){\makebox(0,0)[t]{$K^0$}} 
\put(127,00){\line(1,0){26}} 
\put(160,00){\circle{15}}
\put(160,15){\makebox(0,0)[b]{$V^-$}} 
\put(160,-15){\makebox(0,0)[t]{$K^-$}} 
\end{picture}
\end{minipage}
\ee
$$
\Downarrow
$$
\be
\begin{minipage}{260pt}
\setlength{\unitlength}{1pt}
\small\thicklines
\begin{picture}(260,55)(-10,-30)
\put( 80,00){\circle{15}}
\put( 60,15){\makebox(0,0)[b]{$V^+\wedge (V^0-1)$}} 
\put( 80,-15){\makebox(0,0)[t]{$K^+$}}  
\put( 87,00){\line(1,0){26}} 
\put(120,00){\circle{15}}
\put(115,-5){\line(1, 1){10}}  
\put(115, 5){\line(1,-1){10}}  
\put(120,15){\makebox(0,0)[b]{$-V^0$}} 
\put(120,-15){\makebox(0,0)[t]{$\widetilde{K^0}$}} 
\dottedline{3}(128,0)(152,0) 
\put(160,00){\circle{15}}
\put(180,15){\makebox(0,0)[b]{$V^-\wedge (V^0+1)$}} 
\put(160,-15){\makebox(0,0)[t]{$K^-$}} 
\end{picture}
\end{minipage}
\ee

\bigskip
The wedge product is discussed in the previous papers. In the following, we shall need the following two simple cases only
\be
0\wedge N = N,\qquad\qquad N\wedge (-N) = 0.
\ee

\medskip
{\bf Dualization II}: $V^0 = 0$, \ \ \ $\widetilde{K^0} = K^++K^--K^0-1$
\be
\begin{minipage}{260pt}
\setlength{\unitlength}{1pt}
\small\thicklines
\begin{picture}(260,55)(-10,-30)
\put( 80,00){\circle{15}}
\put( 80,15){\makebox(0,0)[b]{$V^+$}} 
\put( 80,-15){\makebox(0,0)[t]{$K^+$}}  
\dottedline{3}(88,0)(112,0)  
\put(120,00){\circle{15}}
\put(115,-5){\line(1, 1){10}}  
\put(115, 5){\line(1,-1){10}}  
\put(120,15){\makebox(0,0)[b]{$0$}} 
\put(120,-15){\makebox(0,0)[t]{$K^0$}} 
\put(127,00){\line(1,0){26}} 
\put(160,00){\circle{15}}
\put(160,15){\makebox(0,0)[b]{$V^-$}} 
\put(160,-15){\makebox(0,0)[t]{$K^-$}} 
\end{picture}
\end{minipage}
\ee
\be
\Downarrow
\ee
\be
\begin{minipage}{260pt}
\setlength{\unitlength}{1pt}
\small\thicklines
\begin{picture}(260,55)(-10,-30)
\put( 80,00){\circle{15}}
\put( 80,15){\makebox(0,0)[b]{$V^+$}} 
\put( 80,-15){\makebox(0,0)[t]{$K^+$}}  
\put( 87,00){\line(1,0){26}} 
\put(120,00){\circle{15}}
\put(115,-5){\line(1, 1){10}}  
\put(115, 5){\line(1,-1){10}}  
\put(120,15){\makebox(0,0)[b]{$0$}} 
\put(120,-15){\makebox(0,0)[t]{$\widetilde{K^0}$}} 
\dottedline{3}(128,0)(152,0) 
\put(160,00){\circle{15}}
\put(160,15){\makebox(0,0)[b]{$V^-$}} 
\put(160,-15){\makebox(0,0)[t]{$K^-$}} 
\end{picture}
\end{minipage}
\ee

\subsection{Application to the Bethe equations for $\Phi_{\rm gauge}$}

The superconformal primary $\Phi_{\rm gauge}$ has the following excitation numbers
\be
\begin{minipage}{260pt}
\setlength{\unitlength}{1pt}
\small\thicklines
\begin{picture}(260,55)(-10,-30)
\dottedline{3}(-32,0)(-10,0)  
\put(  0,00){\circle{15}}
\dottedline{3}(8,0)(32,0)    
\put( 40,00){\circle{15}}     
\put( 35,-5){\line(1, 1){10}}  
\put( 35, 5){\line(1,-1){10}}  
\put( 40,-15){\makebox(0,0)[t]{$N+2$}} 
\put( 47,00){\line(1,0){26}} 
\put( 80,00){\circle{15}}
\put( 80,-15){\makebox(0,0)[t]{$N+3$}}  
\put( 87,00){\line(1,0){26}} 
\put(120,00){\circle{15}}
\put(120,15){\makebox(0,0)[b]{$+1$}} 
\put(120,-15){\makebox(0,0)[t]{$N+4$}} 
\put(127,00){\line(1,0){26}} 
\put(160,00){\circle{15}}
\put(160,-15){\makebox(0,0)[t]{$N+2$}} 
\put(167,00){\line(1,0){26}} 
\put(200,00){\circle{15}}
\put(195,-5){\line(1, 1){10}} 
\put(195, 5){\line(1,-1){10}} 
\put(200,-15){\makebox(0,0)[t]{$N$}} 
\dottedline{3}(208,0)(232,0) 
\put(240,00){\circle{15}}
\dottedline{3}(250,0)(270,0) 
\end{picture}
\end{minipage}
\ee
Dualizing at 2
\be
\begin{minipage}{260pt}
\setlength{\unitlength}{1pt}
\small\thicklines
\begin{picture}(260,55)(-10,-30)
\dottedline{3}(-32,0)(-10,0)  
\put(  0,00){\circle{15}}
\put( -5,-5){\line(1, 1){10}}  
\put( -5, 5){\line(1,-1){10}}  
\put(  7,00){\line(1,0){26}} 
\put( 40,00){\circle{15}}     
\put( 35,-5){\line(1, 1){10}}  
\put( 35, 5){\line(1,-1){10}}  
\dottedline{3}(48,0)(72,0)   
\put( 80,00){\circle{15}}
\put( 75,-5){\line(1, 1){10}}  
\put( 75, 5){\line(1,-1){10}}  
\put( 80,-15){\makebox(0,0)[t]{$N+3$}}  
\put( 87,00){\line(1,0){26}} 
\put(120,00){\circle{15}}
\put(120,15){\makebox(0,0)[b]{$+1$}} 
\put(120,-15){\makebox(0,0)[t]{$N+4$}} 
\put(127,00){\line(1,0){26}} 
\put(160,00){\circle{15}}
\put(160,-15){\makebox(0,0)[t]{$N+2$}} 
\put(167,00){\line(1,0){26}} 
\put(200,00){\circle{15}}
\put(195,-5){\line(1, 1){10}} 
\put(195, 5){\line(1,-1){10}} 
\put(200,-15){\makebox(0,0)[t]{$N$}} 
\dottedline{3}(208,0)(232,0) 
\put(240,00){\circle{15}}
\dottedline{3}(250,0)(270,0) 
\end{picture}
\end{minipage}
\ee
Dualizing at 3
\be
\begin{minipage}{260pt}
\setlength{\unitlength}{1pt}
\small\thicklines
\begin{picture}(260,55)(-10,-30)
\dottedline{3}(-32,0)(-10,0)  
\put(  0,00){\circle{15}}
\put( -5,-5){\line(1, 1){10}}  
\put( -5, 5){\line(1,-1){10}}  
\put(  7,00){\line(1,0){26}} 
\put( 40,00){\circle{15}}     
\put( 47,00){\line(1,0){26}} 
\put( 80,00){\circle{15}}
\put( 75,-5){\line(1, 1){10}}  
\put( 75, 5){\line(1,-1){10}}  
\dottedline{3}(88,0)(112,0)  
\put(120,00){\circle{15}}
\put(115,-5){\line(1, 1){10}}  
\put(115, 5){\line(1,-1){10}}  
\put(120,15){\makebox(0,0)[b]{$+1$}} 
\put(120,-15){\makebox(0,0)[t]{$N+4$}} 
\put(127,00){\line(1,0){26}} 
\put(160,00){\circle{15}}
\put(160,-15){\makebox(0,0)[t]{$N+2$}} 
\put(167,00){\line(1,0){26}} 
\put(200,00){\circle{15}}
\put(195,-5){\line(1, 1){10}} 
\put(195, 5){\line(1,-1){10}} 
\put(200,-15){\makebox(0,0)[t]{$N$}} 
\dottedline{3}(208,0)(232,0) 
\put(240,00){\circle{15}}
\dottedline{3}(250,0)(270,0) 
\end{picture}
\end{minipage}
\ee
Dualizing at 4, and working at twist-3, $L=3$, we obtain 
\be
\widetilde{K_4} = L+K_3+K_5-K_4-1 = 3 + (N+2)-(N+4)-1 = 0 \ \ !
\ee
\be
\begin{minipage}{260pt}
\setlength{\unitlength}{1pt}
\small\thicklines
\begin{picture}(260,55)(-10,-30)
\dottedline{3}(-32,0)(-10,0)  
\put(  0,00){\circle{15}}
\put( -5,-5){\line(1, 1){10}}  
\put( -5, 5){\line(1,-1){10}}  
\put(  7,00){\line(1,0){26}} 
\put( 40,00){\circle{15}}     
\put( 47,00){\line(1,0){26}} 
\put( 80,00){\circle{15}}
\put( 87,00){\line(1,0){26}} 
\put(120,00){\circle{15}}
\put(115,-5){\line(1, 1){10}}  
\put(115, 5){\line(1,-1){10}}  
\put(120,15){\makebox(0,0)[b]{$-1$}} 
\dottedline{3}(128,0)(152,0) 
\put(160,00){\circle{15}}
\put(155,-5){\line(1, 1){10}}  
\put(155, 5){\line(1,-1){10}}  
\put(160,15){\makebox(0,0)[b]{$+2$}} 
\put(160,-15){\makebox(0,0)[t]{$N+2$}} 
\put(167,00){\line(1,0){26}} 
\put(200,00){\circle{15}}
\put(195,-5){\line(1, 1){10}} 
\put(195, 5){\line(1,-1){10}} 
\put(200,-15){\makebox(0,0)[t]{$N$}} 
\dottedline{3}(208,0)(232,0) 
\put(240,00){\circle{15}}
\dottedline{3}(250,0)(270,0) 
\end{picture}
\end{minipage}
\ee
Dualizing at 5, we find 
\be
\widetilde{K_5} = L+K_4+K_6-K_5-1 = 3 + N - (N+2)-1 = 0 \ \ !
\ee
\be
\begin{minipage}{260pt}
\setlength{\unitlength}{1pt}
\small\thicklines
\begin{picture}(260,55)(-10,-30)
\dottedline{3}(-32,0)(-10,0)  
\put(  0,00){\circle{15}}
\put( -5,-5){\line(1, 1){10}}  
\put( -5, 5){\line(1,-1){10}}  
\put(  7,00){\line(1,0){26}} 
\put( 40,00){\circle{15}}     
\put( 47,00){\line(1,0){26}} 
\put( 80,00){\circle{15}}
\put( 87,00){\line(1,0){26}} 
\put(120,00){\circle{15}}
\put(127,00){\line(1,0){26}} 
\put(160,00){\circle{15}}
\put(155,-5){\line(1, 1){10}}  
\put(155, 5){\line(1,-1){10}}  
\put(160,15){\makebox(0,0)[b]{$-2$}} 
\dottedline{3}(168,0)(192,0) 
\put(200,00){\circle{15}}
\put(200,15){\makebox(0,0)[b]{$+3$}} 
\put(200,-15){\makebox(0,0)[t]{$N$}} 
\dottedline{3}(208,0)(232,0) 
\put(240,00){\circle{15}}
\dottedline{3}(250,0)(270,0) 
\end{picture}
\end{minipage}
\ee
which are the Bethe equations for $XXX_{-s}$ with $s = \frac{3}{2}$.

\section{Perturbative solution of the long-range Bethe equations for $\Phi_{\rm gauge}$}

The long-range (asymptotic) Bethe equations for the full $\mathfrak{psu}(2,2|4)$ theory have been 
proposed in~\cite{Beisert:2005fw}. Unfortunately, they do not have the same large set of duality transformations
that we have discussed for the one-loop equations. Therefore, 
it is non trivial to repeat the reduction to a simple $XXX_{-\frac{3}{2}}$ chain. However, this is not 
our main aim. Instead, we want to obtain a perturbative expansion of the solution associated to the state $\Phi_{\rm gauge}$
which starts from the one-loop solution as an input. This is relatively easy, as we now explain. In principle, there can be 
better methods, but the one we present is rather simple and makes the job.

First, we observe that the long-range Bethe equations have been proposed in 4 equivalent forms. The most convenient one 
has the following degree assignment
\be
\begin{minipage}{260pt}
\setlength{\unitlength}{1pt}
\small\thicklines
\begin{picture}(260,55)(-10,-30)
%
\put(-32,0){\line(1,0){22}}  
\put(  0,00){\circle{15}}
\put( -5,-5){\line(1, 1){10}}  
\put( -5, 5){\line(1,-1){10}}  
\dottedline{3}(8,0)(32,0)    
\put( 40,00){\circle{15}}     
\dottedline{3}(48,0)(72,0)   
\put( 80,00){\circle{15}}
\put( 75,-5){\line(1, 1){10}}  
\put( 75, 5){\line(1,-1){10}}  
\put( 80,-15){\makebox(0,0)[t]{$N+3$}}  
\put( 87,00){\line(1,0){26}} 
\put(120,00){\circle{15}}
\put(120,15){\makebox(0,0)[b]{$+1$}} 
\put(120,-15){\makebox(0,0)[t]{$N+4$}} 
\put(127,00){\line(1,0){26}} 
\put(160,00){\circle{15}}
\put(155,-5){\line(1, 1){10}}  
\put(155, 5){\line(1,-1){10}}  
\put(160,-15){\makebox(0,0)[t]{$N+2$}} 
\dottedline{3}(168,0)(192,0) 
\put(200,00){\circle{15}}
\put(200,-15){\makebox(0,0)[t]{$1$}} 
\dottedline{3}(208,0)(232,0) 
\put(240,00){\circle{15}}
\put(235,-5){\line(1, 1){10}} 
\put(235, 5){\line(1,-1){10}} 
\put(250,0){\line(1,0){20}} 
\end{picture}
\end{minipage}
\ee
The excitation numbers are those of $\Phi_{\rm gauge}$ and are obtained from the Beauty form after dualization 
of nodes 2, 5 followed by dualization of nodes 1, 7.
It can be checked that the single root at node 6 vanishes by symmetry.

There are $3\,N+10$ roots (actually one of them is identically zero) and the solution of the Bethe equations is non trivial even at one loop. However, 
we can make a very useful observation. Along a chain of one-loop duality we can easily backtrace the dualization of the Bethe roots.
This means that we can start from the $N$ Bethe roots of the $XXX_{-3/2}$ chain and compute backward the $3\,N+10$ roots in the above diagram.
This can be done with arbitrarily high precision. The  $XXX_{-3/2}$ roots can be found by solving the Baxter equation and this amounts to finding the roots
of a single polynomial. Each backstep also requires the determination of the roots of a polynomial. All this numerical tasks can be done 
robustly at high precision. 

Once, we have the one-loop solution of the above Bethe equation, it is straightforward to evaluate their perturbative expansion with the 
long-range all-loop deformed version of the equations. The resulting anomalous dimension has rational coefficients in its loop expansions. These
rational numbers can be easily and unambiguously identified according to the methods discussed in~\cite{Beccaria:2007cn,Kotikov:2007cy}.

\section{Three loop anomalous dimensions in the gauge sector}

We believe that the above procedure can be carried out safely at least up to the three loop level. We do not know
if wrapping terms appear at four loops and leave this important issue for future investigations.
Expanding the anomalous dimension  ( we omit $L=3$ and the label $A_\mu$)
\be
\gamma(N) = \sum_{k=1}^\infty g^{2\,k}\,\gamma_k(N),
\ee
we have to reproduce the rational values $\gamma_k(N)$, $k = 1, 2, 3$, by a suitable closed analytical formula. Of course this is not a well-posed problem.
In the case of twist-2 operators or twist-3 scalar and gaugino channels, it has been possible to accomplish the task resorting to the KLOV
principle. Here, already at one-loop, the KLOV principle is violated !

Inspired by other similar QCD calculations~\cite{Mertig:1995ny}, we have made the following Ansatz which generalizes the one-loop result
\be
\label{eq:weakklov}
\gamma_k(N) = \sum_{p=0}^{2\,k+1}\,\sum_{q=0}^{p}\sum_{F_q\in H_q}\, c_{p, F_q}\,\frac{F_q(n)}{(n+1)^{p-q}},\qquad n = \frac{N}{2}+1,
\ee
where $F_q\in H_q$ are linearly independent products of (nested) harmonic sums with positive indices and total transcendentality $q$ all evaluated at 
argument $n$. The terms with $p=2\,k+1$ and $q=p$ are the maximum transcendentality ones. The other terms have subleading transcendentality.

The unknown coefficients in the above Ansatz can be (largely over)determined by computing $\gamma_i(N)$ for a large set of 
spin values. In the end, we arrived at the following remarkable expressions of the two loop anomalous dimension (we rewrite also $\gamma_1$ for completeness)
\ba
\label{eq:gamma2}
\gamma_1 &=& 4\,S_1+\frac{2}{n+1}+4, \nonumber \\
\gamma_2 &=& -2\,S_3-4\,S_1\,S_2-\frac{2\,S_2}{n+1}-\frac{2\,S_1}{(n+1)^2}-\frac{2}{(n+1)^3} + \\
&& -4\,S_2-\frac{2}{(n+1)^2}-8, \nonumber 
\ea
and of the three loop contribution
\ba
\label{eq:gamma3}
\gamma_3 &=& 
5\,S_5
+6\, S_2\, S_3
-4\, S_{2,3}
+4\, S_{4,1}
-8\,S_{3,1,1} \\
&&
+\left(4\, S_2^2+2\,S_4+8\,S_{3,1}\right)\,S_1 \nonumber \\
&&
+\frac{-S_4+4\,S_{2,2}+4\, S_{3,1}}{n+1}
+\frac{4\, S_1\, S_2+S_3}{(n+1)^2}
+\frac{2\, S_1^2+3\, S_2}{(n+1)^3} \nonumber \\
&&
+\frac{6\, S_1}{(n+1)^4}
+\frac{4}{(n+1)^5} 
-2 \,S_4
+8\,S_{2,2} 
+8\, S_{3,1} \nonumber \\
&&
+\frac{4\, S_2}{(n+1)^2}
+\frac{4\, S_1}{(n+1)^3}
+\frac{6}{(n+1)^4} 
+ 8\, S_2
+32, \nonumber
\ea
where $n = \frac{N}{2}+1$ and in all harmonic sums $S_{\bf a}\equiv S_{\bf a}(n)$.

One can identify several pieces which also appeared in the scalar sector. The additional terms have a non-trivial structure that shall be 
further discussed in~\cite{forth}. To give an example, one immediately notice that it is possible to recast the two loop anomalous dimension
in the following compact and symmetric form 
\ba
\gamma_1 &=& 2\,S_1(n)+2\,S_1(n+1)+4, \\
\gamma_2 &=& -2\,S_3(n)-2\,\left[S_1(n)\,S_2(n)+S_1(n+1)\,S_2(n+1)\right] \nonumber\\
&& -2\,\left[S_2(n)+S_2(n+1)\right]-8.\nonumber
\ea
As a non trivial check of Eqs.~(\ref{eq:gamma2},\ref{eq:gamma3}), it is easy to check that the correct three loop 
scaling function is reproduced by the leading large $N$ terms. Expanding $f(g)$ in Eq.~(\ref{eq:scaling})
\be
f(g) = \sum_{n=1}^\infty g^{2\,n} f_n,
\ee
we find the coefficients $f_n$ from the asymptotic values of the maximal transcendentality harmonic combinations 
multiplying $S_1\sim \log N$. These are
\ba
f_1 &=& 4,  \\
f_2 &=& -4\,S_2(\infty),\\
f_3 &=& 4\,S_2^2(\infty)+2\,S_4(\infty)+8\,S_{3,1}(\infty). \nonumber
\ea
Using the exact values
\be
S_2(\infty) = \zeta_2 = \frac{\pi^2}{6}, \quad
S_4(\infty) = \zeta_4 = \frac{\pi^4}{90},\quad
S_{3,1}(\infty) = \frac{\pi^4}{72},
\ee
we recover
\be
f(g)  = 4\,g^2-\frac{2\,\pi^2}{3}\,g^4 + \frac{11\,\pi^4}{45}\,g^6 + \cdots\ .
\ee

\section{A further non-trivial test: MVV-like relations}
\label{sec:MVV}

As a further test of the proposed three loop anomalous dimension, one can check the validity of generalized 
Moch-Vermaseren-Vogt (MVV) relations~\cite{Moch:2004pa,Vogt:2004mw}. These are discussed in full details in 
the recent papers~\cite{Belitsky:2006en,Beccaria:2007bb}.
One assumes that $\gamma(N)$ obeys at all orders the non-linear equation
\be
\label{eq:nonlinear}
\gamma(s) = P\left(N+\frac{1}{2}\gamma(N)\right),
\ee
with a function $P$ admitting the following {\em reciprocity respecting} or {\em parity respecting} expansion for large argument
\be
P(N) = A'\,\log J^2(N) + \sum_{n=0}^\infty\sum_{m=0}^n\,B_{n,m}'\,\frac{\log^m J^2(N)}{(J^2(N))^n}, 
\ee
where the collinear Casimir is 
\be
J^2(N) = (N+L\,s-1)(N+L\,s),\qquad (s = \frac{3}{2},\  L = 3).
\ee
If we now expand $\gamma(N)$ at large $N$ according to 
\ba
\gamma(N) &=& A\,\log\widehat{N} + \sum_{n=0}^\infty\sum_{m=0}^n\,B_{n,m}\,\frac{\log^m\widehat{N}}{N^n}, \\
\widehat{N} &=& \frac{1}{2}\,N\,e^{\gamma_E},
\ea
we can eliminate the coefficients $A'$ and $B'_{n,m}$ and find
all order relations among $A$ and $B_{n,m}$. The first MVV-like relations are 
\ba
\label{eq:mvv}
B_{1,1} &=& \frac{1}{2}\,A^2, \\
B_{1,0} &=& A\,\left(4+\frac{1}{2}\,B_{0,0}\right).\nonumber
\ea
To check them, we compute the large $N$ expansion of $\gamma_{1,2,3}$. It is
\ba
\gamma_1 &=& 4\,\log\widehat{N}+4+\frac{16}{N}-\frac{100}{3\,N^2} + \cdots, \\
\gamma_2 &=& -4\,\zeta_2\,\log\widehat{N}-2\,\zeta_3-4\,\zeta_2-8 \nonumber\\
&& + \frac{8}{N}(\log\widehat{N}-2\,\zeta_2+1) + \frac{4}{N^2}\left(-8\,\log\widehat{N}+\frac{25}{3}\,\zeta_2+1\right) + \cdots \nonumber\\
\gamma_3 &=& \frac{44}{5}\,\zeta_2^2\,\log\widehat{N}-\zeta_5+2\,\zeta_3\,\zeta_5+\frac{44}{5}\,\zeta_2^2+8\,\zeta_2+32\nonumber\\
&& -\frac{4}{5\,N}(20\,\zeta_2\,\log\widehat{N}-44\,\zeta_2^2+5\,\zeta_3+20\,\zeta_2+20)\nonumber\\
&&-\frac{4}{3\,N^2}(6\,\log^2\widehat{N}-48\,\zeta_2\,\log\widehat{N}+55\,\zeta_2^2-12\,\zeta_3+3\,\zeta_2-45) + \cdots . \nonumber
\ea
Hence, the above coefficients of the expansion are
\ba
A &=& 4\,g^2-4\,\zeta_2\,g^4+\frac{44}{5}\,\zeta_2^2\,g^6 + \cdots, \\
B_{0,0} &=& 4\,g^2-(8+4\,\zeta_2+2\,\zeta_3)\,g^4 + \cdots, \\
B_{1,1} &=& 8\,g^4-16\,\zeta_2\,g^6 + \cdots,  \\
B_{1,0} &=& 16\,g^2+8\,(1-2\,\zeta_2)\,g^4-\frac{4}{5}(-44\,\zeta_2^2+20\,\zeta_2+5\,\zeta_3+20)\,g^6 + \cdots,
\ea
and one checks immediately that Eqs.~(\ref{eq:mvv}) hold. Of course, $A$ is nothing but the scaling function $f(g)$.

A detailed analysis of the function $P$ as well as a rigorous proof of its reciprocity properties will appear in a 
forthcoming paper~\cite{forth}.

\section{Conclusions}
\label{sec:conc}

The main result of this paper is the three loop expression of the anomalous dimension $\gamma(N)$ of finite spin $N$ 
maximal helicity twist-3 gluon operators in $\mk{psu}(2,2|4)$ reported in Eqs.~(\ref{eq:gamma2},\ref{eq:gamma3}).
We have obtained them, by solving perturbatively the long-range Bethe equations and resumming the rational expansion
of $\gamma(N)$ assuming the Ansatz Eq.~(\ref{eq:weakklov}).

From the technical point of view, this result exploits the one-loop equivalence of this sector with the integrable 
$XXX_{-3/2}$ spin chain, as follows from a sequence of dualizations of the associated Bethe equations. Beyond one-loop, 
the available duality relations are quite less powerful and should be extended, at least in principle, as discussed
in~\cite{Beisert:2005di}. It would be nice to obtain a reduced set of asymptotic multi-loop Bethe equations of minimal rank.
This interesting task is an open issue that is left for future investigations. 

A more interesting topic concerns the physics encoded in Eqs.~(\ref{eq:gamma2},\ref{eq:gamma3}). From this point of 
view, the fact that in the infinite spin limit we recover the correct cusp anomalous dimension is a mere check definitely not
surprising, but reassuring. On the other hand, the generalized Moch-Vermaseren-Vogt relations discussed in Sec.~(\ref{sec:MVV})
are actually non trivial. They suggest hidden reciprocity relations governing the large spin expansion of $\gamma(N)$.
They hold true for all known results about twist-2 anomalous dimensions in QCD and  ${\cal N}=4$ SYM (even at strong coupling)~\cite{Basso:2006nk,Dokshitzer:2006nm}.
An easy calculation based on the results of~\cite{Beccaria:2007vh} confirms that they are satisfied also in the twist-3 gaugino channel, exploiting the relation
with the twist-2 universal anomalous dimension. Finally, they have been recently checked at four loops in the case of 
twist-3 bosonic $\mk{sl}(2)$ operators  in ${\cal N}=4$ SYM~\cite{Beccaria:2007bb}.
It would certainly be interesting to prove them from first principles at the level of Bethe Ansatz equations.

\acknowledgments
We thank M.~Staudacher for many suggestions and useful comments. We also thank 
G.~Marchesini, Yu.~L.~Dokshitzer, and G.~Korchemsky, for discussions.

\end{document}